\documentstyle[prl,aps,multicol]{revtex}
\begin{document}
\tightenlines
\def\CC{{\rm\kern.24em \vrule width.04em height1.46ex depth-.07ex
\kern-.30em C}}
\def\RR{{\rm
         \vrule width.04em height1.58ex depth-.0ex
         \kern-.04em R}}
\def\ZZ{{\sf Z\kern-.44em Z}}
 \def\id{{\rm 1\kern-.22em l}}

\title{One-Dimensional ${\bf XXZ}$ Model for Particles Obeying 
Fractional Statistics} 
\author{Luigi Amico$^{\dagger,\natural}$, 
Andreas Osterloh$^{\dagger}$, and Ulrich Eckern$^{\dagger}$}
\address{$^{\dagger}$Institut f\"ur Physik,Universit\"at Augsburg,
        D-86135 Augsburg, Germany}
\address{$^{\natural}$ Istituto di Fisica, Facolt\'a di 
Ingegneria, Universit\`a
di Catania $\&$ $INFM$, viale A. Doria 6, I-95129 Catania, Italy}
\maketitle
\begin{abstract}
We define one-dimensional particles as  non-abelian representations 
of the symmetric 
group $S_N$. The exact solution  of an $XXZ$ type 
Hamiltonian built up with such particles
is  achieved using the coordinate Bethe Ansatz. The Bethe equations show
that fractional statistics, effectively, accounts for 
coupling an external gauge 
field to an integer statistics' system.
\end{abstract}

\medskip


\begin{multicols}{2}
\narrowtext
Physical behaviour of quantum systems is deeply affected by the statistics of  
the constituting effective degrees of freedom.
Quasi-particles and quasi-holes  in condensed matter physics 
may obey  statistics interpolating  between fermionic and 
bosonic behaviour. 
Examples are the excitations of two-dimensional electron 
systems exhibiting  Fractional Quantum Hall 
effect~\cite{HALL}. 
These excitations are called {\it anyons}. They  have been a subject of 
intense study also in connection  with superconductivity~\cite{WIL} 
and superfluidity~\cite{VOLOVIK}. Fractional statistics of 
such particles arises from the 
trajectory-dependence of the  particle exchange 
procedure in the two-dimensional 
configuration space.
This feature makes the concept of  anyons  purely two-dimensional. 
The Fock space formulation of anyon operator algebras takes 
into account these characteristics. 
The creation and annihilation operators 
(introduced as Jordan-Wigner transforms of usual fermions
on a two-dimensional lattice\cite{LERDASCIUTO} or as 
unitary representations of the 
diffeomorphism group of $\RR^2$~\cite{GOLDIN}) obey 
deformed commutation relations if 
the exchange involves anyons at different spatial positions (see Appendix).
$N$-anyon-states are abelian representations of the braid 
group $B_{N}$~\cite{LERDA} 
(whereas bosons and fermions furnish, respectively, the identical and 
alternating abelian representations of the symmetric group $S_{N}$).
These features make anyons different from $q$-oscillators,
the latter providing a realization of Gel'fand-Farlie quantum group, 
which is a {\it local} deformation of the Weyl-Heisenberg 
(bosons) or Clifford algebra 
(fermions)~\cite{GELFAND}. 
The path dependence implies that the one-particle state is inextricably related
with the complete state of the many body configuration.
This intrinsic non-locality makes anyon physics very difficult. 
Even statistical properties of a free anyon gas are only partially established using the virial 
expansion~\cite{VIRIAL}.   

Haldane~\cite{HALDANE} formulated  the notion of fractional 
statistics without any reference to the spatial dimension $D$. 
The generalized Pauli principle is expressed in terms of the reduction 
of the single-particle Hilbert space   when particles are added to a many 
body system  keeping  boundary conditions fixed.  
Another way to introduce  dimensionality-independent fractional statistics 
has been formulated in  Ref.~\onlinecite{GREENBERG} where      
{\it quons} have been introduced. Quons' fractional statistics results from the 
``superposition''  of statistical  properties of bosons and fermions~\cite{WU}. 
In $D>2$, this is consistent with spin-statistics theorem. 
In $2D$, Haldane particles and  quons capture the essential features 
of anyons~\cite{HALDANE,GOLDIN}.
   
Recently, an outgrowing interest has been devoted to generalized statistics in one 
dimension.
A specific way to introduce $D=1$ fractional statistics has been proposed in connection
to the quantization of the solutions of the Calogero model~\cite{CALOGERO}. 
There, the potential $1/x^{2}$ is interpreted as ``statistics interaction''. 
The same notion of fractional statistics 
applies also to anyons in a strong magnetic field that 
restricts the allowed energies to the lowest Landau level. The anyon gas, then, is described 
by an effective   field theory on a ring where  
the dynamics of particles is one-dimensional~\cite{1DANY}. It is worthwhile noting that such one-dimensional particles obey  fractional statistics, but they are not ``true anyons'' since in 
$D \neq 2$ trajectories in the particle configuration space have no meaningful braiding property. 
Instead, nonlocal ``deformations'' 
of the commutation relations furnish   {\it non-abelian} representations of the symmetric 
group $S_{N}$.
 
In this paper, we deal with  particles in $D=1$ that 
preserve the intrinsic non-locality of two-dimensional anyons, but which are still representations
of $S_{N}$. This representation is no longer abelian.   
The second quantized  formalism and the Fock space representation 
is developed.
The $XXZ$ model for such particles is formulated and solved exactly  
using coordinate Bethe Ansatz (BA)~\cite{BETHE} in $D=1$.

For $D=1$ we define a set of creation/annihilation operators $\mbox{$\{f_i^{\dagger}\, ,\, f_i\}$}$ 
for a spinless particle at site $i$. They obey the deformed relations

\begin{eqnarray}
f^\dagger_j f_{{ k}} + q_{{ j}, {k}}\,  f_{{ k}} f^\dagger_j &=& 
\delta_{{j}, {k}} \quad ,  
\label{any1} 
\\
f_{ j} f_{ k} + q^{-1}_{j,k} \,  f_{k } f_{ j} &=& 0 \qquad , 
\label{any2} 
\end{eqnarray}
where $q^{-1}_{j,k}\doteq (q_{j,k})^{-1}$.
Since the  operators are path-independent in $D\neq 2$
(compare Appendix),        
(\ref{any1}) and (\ref{any2})  have  to constitute a representation of $S_N$,
and not of $B_N$.
This is ensured by the  ``consistency relations''
\begin{equation} 
q_{{ j}, { k}} = q^{-1}_{{ k}, { j}} = q^{\dagger}_{k,j} \quad , 
\label{CONS1}
\end{equation} 

\begin{equation}
[ {f}_j f^{\dagger}_k , q_{j,k} ] =0 \quad .
\label{CONS2}
\end{equation}
Such a representation is non-abelian~\cite{LERDA}. 
For $j=k$, (\ref{CONS1}) gives $q_{j,j}=\pm\id $.
Hence (\ref{any1}), (\ref{any2}) are an extension of anyon commutation 
relations~\cite{LERDASCIUTO,GOLDIN} to $D \neq 2 $.
In contrast to the true anyonic case (see Appendix), $q_{j,k}$ has no 
relation with the configuration space geometry,  but is a free 
``external'' parameter.  

Relations (\ref{any1}) and (\ref{any2}) are formally analog to quon 
commutation rules~\cite{WU}. 
Note that the deformation parameter here depends on two indices $(j,k)$, 
whereas it does not in quon commutation rules.
Without this index-dependence, relation (\ref{CONS1}) directly implies 
$q^2=\id$. As a consequence, quons obey integer statistics in $1D$ if
$q$ is a $\CC$-number~\cite{WU} (as an operator, it has eigenvalues $\pm 1$). 
If $\mbox{$q_{j,k}=\pm \id \quad \forall \ (j,k)$}$, then  (\ref{any1}), 
(\ref{any2}) describe spinless fermions or bosons, respectively. 
However, for application we choose $q_{j,k}$ being $\CC$-numbers and 
$q_{j,j} = 1 $, see (\ref{our-q}), which implies the Pauli exclusion 
principle, as for spinless electrons or hard core bosons.
 
Relations (\ref{CONS1}), (\ref{CONS2}) hold if $q_{j,k}$ is an 
operator commuting or anticommuting with both, ${f}_j$ and $f^{\dagger}_k $. 
For this reason we add this as a postulate~\cite{NOTECOMMUTE}

\begin{equation}\label{postulate-rel} 
[ f^\dagger_{k}\, ,\, q_{j,k}] \ =\ \left [ f_j \, ,\, q_{j,k}\right ] \ = \ 0 \quad .
\end{equation}
The introduction of two indices for the deformation parameter  
allows the construction of consistent commutation relations even for $q_{j,k}$ 
being $\CC$--numbers. We make use of this possibility in (\ref{our-q}). 

To develop a Fock representation of the algebra  (\ref{any1}), (\ref{any2}), 
we take $\nu_{{ j}}\doteq f^\dagger_j f_{j}$ as number operators. 
Relations (\ref{CONS1}), (\ref{CONS2}), and (\ref{postulate-rel}) yield commutators of $\nu_{{ j}}$ and   $f^{\dagger}_{j}$, $f_j$ being unaffected by the deformation parameter $q_{j,k}$: 

\begin{equation}\label{NUMBERCOMM}
[ \nu_{j}, \nu_{k} ]=0 \: ,\;
[ \nu_{j}, f^{\dagger}_{k} ]=
\delta_{j,k} f^\dagger_{k} \: ,\;
[ \nu_{j}, f_{k}]=-
\delta_{j,k} f_{k}\;. 
\end{equation}
Moreover, the property $q_{j,j}=\id $ implies that number operators are idempotent: 
$(\nu_{j})^2=\nu_{j}$.
Because of (\ref{NUMBERCOMM})  the one-particle Fock representation of the algebra 
(\ref{any1}), (\ref{any2})  is unaffected by $q_{j,k}$.
Instead, the action of $ f_l, \, f_l^\dagger, \, \nu_l $ on the $N$-particle state 
$|n_{1}\dots n_{N}\rangle$ is deformed according to  
\begin{eqnarray}
f_l &|n_{1} .. n_{N}\rangle = & {\scriptstyle (-)^{l-1}\delta_{n_l,1}} 
\left [ \prod_{k=1}^{l-1} q_{l,k} \right ] 
|n_{1}..n_l-1..n_{N}\rangle\ ,  \nonumber \\
f_l^{\dagger} &|n_{1} .. n_{N}\rangle = & {\scriptstyle (-)^{l-1} 
\delta_{n_l,0}} 
\left [ \prod_{k=1}^{l-1} q_{k,l} \right ]  
|n_{1}..n_l+1..n_{N}\rangle \ , \nonumber \\
\nu_l &|n_{1} .. n_{N}\rangle = & n_l\, |n_{1}.. n_{N}\rangle \ ,
\label{FOCKREPRE}
\end{eqnarray} 
where $n_l\in \{ 0,1\}$. Equations (\ref{FOCKREPRE}) generalize the 
corresponding relations fulfilled by integer statistics particles~\cite{NEGELE}
characterized by $\prod_{k=1}^{l-1}  q_{l,k}=(\pm)^{l-1}$ (for fermions/bosons).  
An explicit realization of the operators $f_{j}$ in terms of spinless fermionic operators $a_{j}$ is 
$\mbox{$ f_{{ j}}\doteq {a_{j}}\,\exp\left (- i \sum_{ l}\Phi_{ l} n_{{l}} \right ) $}$,
where $\Phi_{l}$ are hermitean operators commuting with fermionic 
degrees of freedom. By direct calculation, the relations (\ref{any1}), (\ref{any2})  are obtained by
setting $\mbox{$ q_{jk}\doteq\exp{\left [i(\Phi_k-\Phi_j )\right ]}$}$. This realization has been suggested in the 
Ref.~\onlinecite{MORA} where $\Phi_{k}\equiv p_{k}$, $p_{k}$ being momenta of a 
phononic bath coupled to fermionic degrees of freedom.

In the following we consider the 1$D$ anisotropic Heisenberg 
model ($XXZ$ model) of spinless fermions

\begin{equation}
H_{XXZ} = - {t}\, \sum_{i } \, \left ( f^\dagger_{i} f_{{i+1}} + 
f^\dagger_{i+1} f_{{i}} \right ) +  U\, \sum_{ i}\nu_{i} \nu_{i+1} \; . 
\label{DEFXXZ}
\end{equation}
The $f$-operators obey relations (\ref{any1}), (\ref{any2}) with $q_{j,k}$ defined in close analogy with anyonic relations (see Appendix)

\begin{eqnarray}\label{our-q}
q_{j,k}=\left \{ \begin{array}{ll} q \quad & j>k  \\
1 \quad & j=k \\
q^{-1} \quad & j<k  \end{array} \right. \, , \, \quad q\in\CC\ \, . 
\end{eqnarray}
We point out that the postulate (\ref{postulate-rel}) is fulfilled since 
$q_{j,k}$ is  a $\CC$--number for arbitrary, fixed $(j,k)$. 
Relations (\ref{CONS1}) and (\ref{our-q}) imply that 
$q$ is on the unit circle, {\it i.e.} $|q|=1$.
Additionally, Periodic Boundary Conditions (PBC) 
$f_{i+L} \equiv f_i $ are chosen, where $L$ denotes the period. The parameters $t$ and $U$ are the 
hopping amplitude and the Coulomb interaction strength, respectively.

It is worthwhile mentioning that (\ref{our-q}) implies fixing an order
on the infinite periodic chain. 
This order can  only be defined locally on the manifold $S^1$, 
which needs two charts for its description.
We choose two ``charts'' $C_1\doteq \{j_1,..,j_N\}$ and 
$C_2\doteq\{j_{N-l},..,j_{N},j_1,..,j_{N-l-1}\}$ on the ring 
thought as a discrete subset of $S^1$. 
On each chart, the given order is well defined by interpreting them as ordered sets.
The intersections between $C_1$ and $C_2$ are $\{j_{N-l}\dots j_N\}$ and $\{j_1\dots j_{N-l-1}\}$.
In such sets, the orders defined on $C_1$, $C_2$ are identical.
Now, in $D=1$, only nearest neighbour ($n.n.$) exchanges can take place. 
Thus $q_{j_k,j_l}$ is connected to the $n.n.$ exchange $j_k\leftrightarrow j_l$.
On the chart $C_1$, where $j_1<j_N$, the exchange $j_N\leftrightarrow j_1$ is not a 
$n.n.$ exchange.
To allow for $n.n.$ hopping $j_N\leftrightarrow j_1$, we  must use $C_2$ on which $j_N<j_1$. 
This implies $q_{j_N,j_1}=q^{-1}$. 
The picture depicted above is equivalent to fixing a period 
$P_0 \doteq \{\,j_1\, ,\,  \dots\, ,\, j_1 + L\, \}$ on the infinite periodic chain.
Consistency of the PBC with this induced order is given if the results
are independent of $P_0$. In the sequel it will be seen that this condition 
is fulfilled. 

The correspondence between   
(\ref{DEFXXZ}) and the deformed anisotropic Heisenberg model can be 
established by $\mbox{$S_j^{(+)} \doteq f^{\dagger}_j$}$, $\mbox{$S_j^{(-)}\doteq f_j$}$ and 
$\mbox{$ S_j^{(z)}\doteq 1/2 -\nu_{j}$}$. On-site, the operators 
$S_j^{(z)}, S_j^{(\pm)}$ generate the fundamental representation 
(spin $s=1/2$) of  $su(2)$, but for $j \neq k$

\begin{eqnarray}
\left [ S_j^{(+)}, S_k^{(-)} \right ]& =&
 (1+q_{j,k}) \; S_k^{(+)}\, S_j^{(-)} -\delta_{j,k} \ ,\nonumber \\
\left [ S_j^{(-)}, S_k^{(-)} \right ]& =&
 (1+q_{j,k}) \; S_k^{(-)}\, S_j^{(-)} -\delta_{j,k} \ ,\nonumber \\
\left [ S_j^{(z)}, S_k^{(\pm )} \right ]& =& 0 \ .
\end{eqnarray} 
We now show that the $XXZ$ model (\ref{DEFXXZ}) 
is exactly solvable by means of the coordinate BA.
The general $N$-particle state on a chain with $L$ sites can be written as

\begin{equation}
|\Psi \rangle \doteq \sum_{1 \le j_1 < ... < j_N \le L}
 \psi \left (  j_1, j_2,... ,j_N \right )\;  
f_{j_{1}}^{\, \dagger}f_{j_{2}}^{\, \dagger}
...f_{j_{N}}^{\, \dagger}|0 \rangle \, .
\end{equation}
The action of $H_{XXZ}$ on $|\Psi \rangle $\/  , {\it i.e.} the eigenvalue equation, then reads

\begin{eqnarray}
&U&\sum_{l=1}^{N-1}\delta_{j_l+1,j_{l+1}} \psi \left (j_1,\dots ,j_N\right )\nonumber \\
&-&t \sum_{l=1}^{N} Q_{l}^{(+)} \psi \left (j_1..j_l -1..j_N\right )
+  Q_{l}^{(-)} \psi \left (j_1..j_l+1..j_N\right )\nonumber \\& &
= E\, \psi \left (j_1,\dots ,j_N\right )\quad ,
\end{eqnarray}
where  
$
Q_l^{(\pm)} \doteq  
\prod_{s=1}^{l-1} q_{j_l, j_s} q_{j_s,j_l \mp 1}
$.
Using (\ref{our-q}) as definition for $q_{j,k}$, it follows that  
\begin{equation}
Q_{l}^{(-)} = Q_{l}^{(+)} = 1\; .
\end{equation}
Hence, exactly the same eigenvalue equation as in the ordinary $XXZ$ model is obtained. 
Consequently, inserting the BA

\begin{equation} 
\psi(j_1,...j_N)=\sum_{\pi \in {\it S}_N} A(\pi)\exp\left (i\sum_{m=1}^N j_{m}k_{\pi(m)}\right )\quad ,
\end{equation}
the energy $E$ and the two--body scattering matrix 
$S(k\, ,\, k') \doteq -\exp{[-i\theta( k \, , \, k')]}$  
for the ``deformed'' model are identical with the known terms occuring in the usual $XXZ$ model~\cite{YANGYANG}.
Imposing PBC, however, yields

\begin{eqnarray}
\psi \left ( j_2,\dots , j_{N}, j_1\right ) &=& \left [\prod_{l=2}^{N} (- q_{j_1,j_l}) \right ] 
\psi \left ( j_{1},j_{2},\dots , j_{N}\right ) \nonumber \\
&=&\ (-q)^{N-1}\, \psi \left ( j_{1},j_{2},\dots , j_{N}\right ) \quad .
\label{TWISTPBC}
\end{eqnarray}
Equation (\ref{TWISTPBC}) shows that the fractional statistics produces a twist in the 
PBC that modifies the periodicity of the Bethe wave function.

Since the twisting factor $q^{N-1}$ does
not depend on  $j_1$, the starting point of the chosen period $P_0$, the boundary condition is consistent with our choice of $q_{j,k}$.
The twist $q^{N-1}$ does not depend on the particles' configuration, but on the number of particles only. 
This is crucial because configuration-dependence of the twisting factor would destroy 
the solvability of the model,  since 
it modifies the structure of the exponential functions in imposing 
PBC on the BA wave function (making it impossible to extract 
from (\ref{TWISTPBC}) a relation for the amplitudes $A(\pi)$). 
So, the coordinate BA solvability of the model (with PBC) 
demands a careful choice of the functional form of $q_{j,k}$.
The choice $q_{j,k}\doteq \exp{[i \delta \, (j-k)]}$ for the $XXZ$ model,
for instance, leads to the same structure of the $S$ matrix, but produces incompatible boundary conditions. 

Since $|q|=1$, $q=\exp{[i\arg(q)]}$. So  
a phase shift by multiples of $\arg(q)$ occurs in the BA wave function 
on the {\it r.h.s.} of eq.~(\ref{TWISTPBC}). The Bethe Equations (BE) are obtained as 

\begin{equation}
  k_j L\ =\  \arg([-q]^{N-1})\  +\ 2\pi I_j\ -\  \sum_{m=1}^{N}\theta ( k_j\, , k_m  ) \ ,
\label{BETHEEQ}
\end{equation}
where $I_j \in \ZZ$. In the fermionic case, one obtains
$k_j L=2\pi (I_j+\frac{N-1}{2})-\sum_{m=1}^{N}\theta ( k_j\, , k_m  )$,
whereas the hard core bosonic case yields $k_j L=2\pi I_j-\sum_{m=1}^{N}\theta ( k_j\, , k_m  )$. In all cases, $I_j \in \ZZ$.
Equation~(\ref{BETHEEQ}) differs from the BE for the ordinary 
$XXZ$ model in the additive term $\arg([-q]^{N-1})$. 
This term has its origin in the fractional statistics of the particles, and vanishes for integer statistics.
Equations (\ref{BETHEEQ}) were obtained for the $1D$ $XXZ$ model on a ring 
threaded by an external magnetic flux~\cite{BATCHELOR,SHASTRY}. 
We recover the BE of \onlinecite{SHASTRY} identifying 
$\Phi\equiv \alpha \doteq \arg([-q]^{N-1}) $, $\Phi$ being the
magnetic flux in units of $\hbar c/ e$. 

The limit $k_N\rightarrow \pi-\mu$, where $\cos(\mu)\doteq U/2t$ for $|U|\leq 2|t|$,  
in the $N$-th equation of~(\ref{BETHEEQ}) 
relates the statistics with the ratio $U/2t$ for the ground state,
\begin{equation}
\alpha= \pi \left ( L-N-2I_N+1  \right ) -\mu \left ( L-2N+2 \right ) \; .  
\end{equation} 
At half filling ($N/L=1/2$, $I_N=(N-1)/2$) the energy and the 
total momentum, $\mbox{$ P\doteq \sum_{l=1}^N k_l/L $}$, of 
the ground state are  
affected by the statistics' factor $\alpha$~\cite{SHASTRY,YANGYANG} as follows: 
\begin{eqnarray}
E_0 (\alpha)- E_{0}(0)&=& {{\pi \sin( \mu)}\over{4\mu(\pi- \mu)L}}\; \alpha^2 \ ,\nonumber \\
P&=&\frac{1}{2} \; \alpha \ ,  
\label{GROUND}
\end{eqnarray}   
where $E_{0}(0)$ denotes the ground state energy of the undeformed $XXZ$ model.  
The same structure of  BE has been obtained in~\onlinecite{ZVYAGIN,SCHULZ}. 
In~\onlinecite{SCHULZ}, the two species of particles, up-spin 
($ \sigma=+ $) and down-spin ($\sigma = - $), have a dynamics 
governed by two distinct $XXZ^{({\sigma})}$ Hamiltonians coupled only via a 
local gauge field, included in $XXZ^{(\sigma )}$ by a Peierls-like substitution
$t\rightarrow t \,  W^{({\sigma})}_{m}$,  where 
$W^{({\sigma})}_{m}\doteq \exp{ [i \sigma \sum_{l=1}^L \alpha_{m-l}\,n_{l,-\sigma} ]}$
(determined by the position of all particles of opposite species); $\alpha_l \in \RR$, 
$\alpha_{m+L}\equiv \alpha_{m}$. 
A comparison of the BE in~\onlinecite{SCHULZ} with (\ref{BETHEEQ}) shows that 
our deformation parameter $q$ can be interpreted as the ``global''  coupling constant 
of the gauge potential by setting  $\alpha \sim \sum_{m=1}^L\alpha_{m-l}$. 
Vice versa, such an interaction produces statistics transmutation. 
In this sense, our deformed $XXZ$ model belongs to the same class of 
integrable models introduced in~\onlinecite{SCHULZ}.  

In conclusion, we have given a  formulation of fractional statistics 
in one dimension realized by  an anyonic-type  deformation 
of the second quantized commutation rules. 
Coordinate BA solvability of the deformed $XXZ$ 
model demands  a proper choice of 
the functional form  of $q_{j,k}$.
The statistics we have chosen in the present paper preserves 
the Yang-Baxter equation as well as the BA 
solvability of the undeformed model.  
The resulting BE are, however, modified. They show that 
fractional statistics plays the 
same role as a gauge field coupled to the undeformed model.
Systematic investigations of  fractional 
statistics seem interesting for at least 
two reasons.
First, fractional statistics may be an alternative 
approach to handle  complicated 
interactions between particles obeying integer statistics. 
Such interactions could be 
modelled  deforming the particles' statistics.
Second, the study of ``compatible statistics'' 
could be relevant in order to  
find  integrable Hamiltonians characterized 
by {\it braided} Yang--Baxter Equations (YBE)~\cite{KUNDU}. 
Such a feature of the YBE 
could be closely related to actual braiding of particles in two dimensions.  

A further development of the present approach is to take spin into account.
A preliminary analysis of the ``deformed'' Hubbard model~\cite{MORA} shows 
that  fractional statistics modifies the $S$ matrix; the $R$ matrix obeys 
a braided YBE. 
We will report on this subject in a forthcoming paper.

\acknowledgments

We thank M. Rasetti 
for suggesting this line of research  as well as for many discussions.
We thank D. Bormann, R. Fazio, G. Falci, G. Giaquinta, P. Schwab, and M. Takahashi for 
fruitful discussions. 
L.A. acknowledges financial support of ``Fondazione A. della Riccia'', 
and EU TMR Programme (Contract no. FMRX CT 960042). This paper has been written 
while the authors were members of
the {\sl Graduiertenkolleg} ``Nonlinear Problems in
Analysis, Geometry, and Physics" (GRK 283)
financed by the German Science Foundation (DFG) and the State of Bavaria.

\begin{appendix}

\vspace*{-0.5cm}
\section{}
Here we summarize the commutation properties of two dimensional 
anyons~\cite{LERDASCIUTO,GOLDIN}.  The creation/annihilation operators  obey 

\begin{eqnarray}
 b^\dagger ({\bf x}_C)\, b ({\bf y}_C) +\; q({\bf x}_{C},{\bf y}_{C}) b 
({\bf y}_C) \, b^\dagger({\bf x}_C) &=& \delta _{{\bf x}_C,{\bf y}_C}\ ,  
\nonumber \\
{b({\bf x}_C)} \, b ({\bf y}_C) + q^{-1}({\bf x}_{C},{\bf y}_{C}) \,  b ({\bf y}_C) \, {b} ({\bf x}_C) &=&0  \;.
\end{eqnarray}
The operators ${b^{\dagger}({\bf x}_C)}$ (${ b({\bf x}_C)}$)  create 
(annihilate) an anyon
at site ${\bf x_C}\doteq (x_1, x_2)$. $C$ denotes 	
the  path running from $+\infty$ to ${\bf x_C}$ keeping $x_2$ constant.
The relations above hold if ${\bf x}_{C} > {\bf y}_{C}$; 
in the case ${\bf x}_{C} < {\bf y}_{C}$,	
they  are satisfied substituting $q \leftrightarrow q^{-1}$. Note that  
${\bf x}_C > {\bf y}_C \Leftrightarrow \{ x_2 > y_2 \;  \vee \;   
x_1> y_1 \; (\hbox{if} \; x_2=y_2) \} $. 
The function  $q({\bf x}_{C},{\bf y}_{C}) = q(|{\bf x}_C-{\bf y}_C|)$ can be simplified 
(see e.g.~\onlinecite{LERDASCIUTO}) 
to $q= e^{i\pi\nu}$ ($\nu\in\RR $), where $\nu$ denotes the {\it statistics}.
If two anyons are at the same position 
${\bf x}_{C} = {\bf y}_{C}$, then  $q=1$. 
Otherwise, the standard bosonic or fermionic algebras are deformed by the 
parameter $q$. 

\end{appendix}

\end{multicols}

\begin{thebibliography}{100}
\bibitem{NOTECOMMUTE} We note that relation (\ref{postulate-rel}) is sufficient 
but not necessary for relation (\ref{CONS2}), but it is 
crucial to obtain (\ref{NUMBERCOMM}).
\end{thebibliography}
\end{document}